\documentclass[aps,prd,twocolumn,nofootinbib,,superscriptaddress,showkeys]{revtex4}
\usepackage{amssymb,amsmath,graphicx,enumerate}
\usepackage{subfigure}

\usepackage{dcolumn,booktabs}
\usepackage{longtable}
\usepackage[dvipdfm,colorlinks=true,citecolor=blue,pdfstartview=FitH]{hyperref}

\def\bea{\begin{equation}}
\def\eea{\end{equation}}
\newcommand{\rt}{Regge trajectory}
\newcommand{\rts}{Regge trajectories}
\newcommand{\bfr}{{\bf r}}
\newcommand{\bfp}{{\bf p}}
\newcommand{\bfpa}{{|\bf p|}}
\newcommand{\gev}{{\rm GeV}}

\newcommand{\sse}{spinless Salpeter equation}

\begin{document}
\title{Regge trajectory relation for the universal description of the heavy-heavy systems: diquarks, mesons, baryons and tetraquarks}
\author{Jiao-Kai Chen}
\email{chenjk@sxnu.edu.cn, chenjkphy@outlook.com }
\affiliation{School of Physics and Information Engineering, Shanxi Normal University, Taiyuan 030031, China}

\begin{abstract}
By employing the nonlinear Regge trajectory relation $M=m_R+\beta_x(x+c_{0x})^{2/3}\,\,(x=l,\,n_r)$, we investigate the heavy-heavy systems, such as the doubly heavy diquarks, the doubly heavy mesons, the heavy-heavy baryons, and the heavy-heavy tetraquarks. The fitted Regge trajectories illustrate that these heavy-heavy systems satisfy the above formula and show the existence of an universal description of the heavy-heavy systems. The universality embodies not only the universal behavior $M{\sim}x^{2/3}$ but also the universal parameters. The values of $c_{fn_r}$ and $c_{fl}$ vary with different heavy-heavy systems, but they are close to one. There is an inequality $\beta_{n_r}>\beta_{l}$, and it holds for all the discussed heavy-heavy systems. Moreover, the expression of $\beta_x$ [Eq. (\ref{parabm})] explains its variation with the change of the constituents' masses.
\end{abstract}


\maketitle


\section{Introduction}
The {\rt} is one of the effective approaches for studying hadron spectra \cite{Regge:1959mz,Collins:1971ff,Collins:1977jy,Inopin:2001ub,
Inopin:1999nf,Chew:1961ev,Chew:1962eu,Irving:1977ea,Wilczek:2004im}. The light hadron spectroscopy has a surprising feature, which is the remarkable similarity between meson and baryon {\rts} \cite{Klempt:2012fy}. In Refs. \cite{Dosch:2015nwa,Brodsky:2018vyy,Nielsen:2018uyn}, the authors provide an unified Regge spectroscopy of mesons, baryons and tetraquarks of the same parity and twist with an universal Regge slope based on the eigensolutions of superconformal algebra.
In Ref. \cite{Sonnenschein:2018fph}, the authors present the masses and widths of predicted excited states of the Holography inspired stringy hadron model.
In Ref. \cite{Chen:2017fcs}, the authors give universal descriptions of the orbitally excited heavy-light mesons and baryons.
In Ref. \cite{Brisudova:1999ut}, a square-root {\rt} is used as an universal form to discuss different types of mesons.
In Ref. \cite{Chen:2023web}, we investigate the universal descriptions of the radially and orbitally excited heavy-light diquarks, mesons, baryons and tetraquarks.
In Refs. \cite{Feng:2023txx,Chen:2023cws,Chen:2023ngj}, the spectra of diquarks and mesons are described universally by employing the {\rt} approach.
It is interesting to investigate whether there exists an universal description of the heavy-heavy systems including the doubly heavy diquarks composed of two heavy quarks, the doubly heavy mesons constituted of one heavy quark and one heavy antiquark, the heavy-heavy baryons composed of a heavy quark and a doubly heavy diquark or a heavy-light diquark, and the heavy-heavy tetraquarks consisting of one doubly heavy diquark or one heavy-light diquark and one doubly heavy antidiquark or one heavy-light antidiquark.

The form of the {\rt} varies with energy regions for different systems \cite{Chen:2021kfw,Chen:2022flh}. For the light systems in which the constituents move relativistically, the {\rt} takes the linear form in the $(M^2,\,x)$ plane, where $M$ is the mass of the bound state. $x=l,n_r$, where $l$ is the orbital angular momentum and $n_r$ is the radial quantum number. For the heavy-heavy systems in which the constituents move nonrelativistically, the {\rt} takes the nonlinear form in the $(M^2,\,x)$ plane. In Ref. \cite{Chen:2018hnx}, it is illustrated that masses of the heavy quarkonia can be well described by the {\rt} relation
\bea\label{ortr}
M^2=\beta(c_ll+{\pi}n_r+c_0)^{2/3}+c_1.
\eea
$\beta$, $c_l$, $c_0$ and $c_1$ are parameters. In Ref. \cite{Feng:2023txx}, it is shown that the doubly heavy mesons and the doubly heavy diquarks can be described universally by employing the {\rt} relation\footnote{The {\rts} of hadrons are commonly plotted in the $(M^2,\,x)$ plane or in the $(x,\,M^2)$ plane, where $x=l,\,n_r$. For simplicity, the figures plotted in the $(M,\,x)$ plane and in the $((M-m_R)^2,\,x)$ plane are also called the {\rts}. In this work, we concentrate on the $\lambda$-mode of baryons and tetraquarks and the $\rho$-mode excitations of diquarks are not considered.}
\bea\label{massform}
M=m_R+\beta_x(x+c_{0x})^{2/3}\,\,(x=l,\,n_r),
\eea
where $m_R$ and $\beta_x$ are in Eq. (\ref{parabm}). Formula (\ref{massform}) is consistent with (\ref{ortr}) by neglecting the small terms.
In this study, we show that the {\rt} relation (\ref{massform}) is appropriate not only for the doubly heavy mesons and the doubly heavy diquarks, but also for other heavy-heavy systems such as the heavy-heavy baryons and the heavy-heavy tetraquarks.

This paper is organized as follows: In Sec. \ref{sec:rgr}, the universal {\rt} relation is obtained from the spinless Salpeter equation (SSE). In Sec. \ref{sec:fit}, the universality of the given {\rt} relation is illustrated by fitting different types of the heavy-heavy systems. In Sec. \ref{sec:disc}, the discussions are presented, and in Sec. \ref{sec:conc}, the conclusions are provided.

\section{{\rt} relations}\label{sec:rgr}
In this section, we present the {\rt} relation obtained from the SSE, which is universal for the heavy-heavy systems.

\subsection{SSE}
The {\sse} (SSE) \cite{Godfrey:1985xj,Ferretti:2019zyh,Durand:1981my,Durand:1983bg,Lichtenberg:1982jp,Jacobs:1986gv,Bedolla:2019zwg} reads as
\begin{eqnarray}\label{qsse}
M\Psi_{d,m,b,t}({\bfr})=M_0\Psi_{d,m,b,t}({\bfr})
+V_{d,m,b,t}\Psi_{d,m,b,t}({\bfr}),
\end{eqnarray}
where $M_0=\omega_1+\omega_2$. $M$ is the bound state mass (diquark, meson, baryon or tetraquark). $\Psi_{d,m,b,t}({\bfr})$ are the diquark wave function, the meson wave function, the baryon wave function and the tetraquark wave function, respectively. $\omega_i$ is the square-root operator of the relativistic kinetic energy of constituent
\bea\label{omega}
\omega_i=\sqrt{m_i^2+{\bf p}^2}=\sqrt{m_i^2-\Delta},
\eea
where $m_1$ is the effective mass of the constituent $1$ (quark or diquark) and $m_2$ is the effective mass of the constituent 2 (quark, antiquark, diquark or antidiquark), respectively.
Even though diquarks are colored states and not physical, diquarks are treated here on an equal footing to mesons, baryons and tetraquarks \cite{Bedolla:2019zwg} as we consider the masses of these states.
Following Ref. \cite{Ferretti:2019zyh,Bedolla:2019zwg}, we employ the potential  \cite{Eichten:1974af,Ferretti:2019zyh,Ferretti:2011zz,Bedolla:2019zwg},
\bea\label{potv}
V_{d,m,b,t}=-\frac{3}{4}\left[V_c+{\sigma}r+C\right]
({\bf{F}}_i\cdot{\bf{F}}_j)_{d,m,b,t},
\eea
where $V_c\,{\propto}\,1/r$ is the color Coulomb potential or a Coulomb-like interaction \cite{Ferretti:2019zyh,Ferretti:2011zz}. The second term is the linear confining potential and $\sigma$ is the string tension. $C$ is a fundamental parameter \cite{Gromes:1981cb,Lucha:1991vn}. ${\bf{F}}_i\cdot{\bf{F}}_j$ is the color-Casmir \cite{Ferretti:2019zyh},
\bea
\langle({\bf{F}}_i\cdot{\bf{F}}_j)_{d}\rangle=-\frac{2}{3},\quad
\langle({\bf{F}}_i\cdot{\bf{F}}_j)_{m,b,t}\rangle=-\frac{4}{3}.
\eea

From Eqs. (\ref{qsse}) and (\ref{potv}), we see that the doubly heavy diquarks, the doubly heavy mesons, the heavy-heavy baryons, and the heavy-heavy tetraquarks are described in an unified approach. Therefore, it is expected that these heavy-heavy systems can be described universally by the {\rt} approach.

\subsection{{\rt} relations}
For the heavy-heavy systems, $m_{1},m_2{\gg}{\bfpa}$, Eq. (\ref{qsse}) reduces to
\begin{eqnarray}\label{qssenr}
M\Psi_{d,m,b,t}({\bfr})&=&\left[(m_1+m_2)+\frac{{\bfp}^2}{2\mu}\right]\Psi_{d,m,b,t}({\bfr})\nonumber\\
&&+V_{d,m,b,t}\Psi_{d,m,b,t}({\bfr}),
\end{eqnarray}
where
\bea
\mu=m_1m_2/(m_1+m_2).
\eea
Following Ref. \cite{Brau:2000st} to employ the Bohr-Sommerfeld quantization approach \cite{brsom} to discuss Eqs. (\ref{potv}) and (\ref{qssenr}) gives
\begin{align}\label{rtnrs1}
M{\sim}&\frac{3}{2}\left(\frac{\sigma'^2}{\mu}\right)^{1/3}l^{2/3} \quad (l{\gg}n_r),\nonumber\\
M{\sim}&\left(\frac{3\pi}{2}\right)^{2/3}\left(\frac{\sigma'^2}{2\mu}\right)^{1/3}{n_r}^{2/3}\quad (n_r{\gg}l),
\end{align}
where
\bea
\sigma'=\left\{\begin{array}{cc}
\sigma/2, & \text{diquarks}, \\
\sigma, & \text{mesons, baryons, tetraquarks}.
\end{array}\right.
\eea
Consider the constant terms and the omitted terms and use Eq. (\ref{rtnrs1}), then we obtain (\ref{massform}) with \cite{Chen:2022flh}
\bea\label{parabm}
\beta_x=c_{fx}c_xc_c,\quad m_R=m_1+m_2+C',
\eea
where
\bea
C'=\left\{\begin{array}{cc}
C/2, & \text{diquarks}, \\
C, & \text{mesons, baryons, tetraquarks}.
\end{array}\right.
\eea
$c_{x}$ and $c_c$ are
\bea\label{cxcons}
c_c=\left(\frac{\sigma'^2}{\mu}\right)^{1/3},\quad c_l=\frac{3}{2},\quad c_{n_r}=\frac{\left(3\pi\right)^{2/3}}{2}.
\eea
Both $c_{fl}$ and $c_{fn_r}$ are equal theoretically to one and are fitted in practice.
In Eq. (\ref{massform}), $m_1$, $m_2$, $\epsilon_c$, $c_x$, $c_{fx}$ and $\sigma$ are universal for the heavy-heavy systems. $c_{0x}$ is universal for points on a given {\rt} and is determined by fitting one point on the given {\rt}.
The simple formula (\ref{massform}) with the coefficients (\ref{cxcons}) can give good results which are consistent with the experimental and theoretical data, see Sec. \ref{sec:fit}.
The {\rt} formula (\ref{massform}) with (\ref{parabm}) obtained from the SSE is not only consistent with the
{\rt} obtained from other potential theories, for example, the Schr\"{o}dinger equation but also consistent with the results obtained from the string theories, for example, the holography inspired stringy hadron model \cite{Sonnenschein:2018fph}. See Refs. \cite{Chen:2021kfw,Chen:2022flh} and references therein for more discussions.
Eq. (\ref{massform}) can be rewritten as another form \cite{Chen:2022flh}
\bea\label{reglike}
(M-m_R)^2=\alpha_x(x+c_{0x})^{4/3}\quad (x=l,\,n_r),
\eea
where $\alpha_x=c^2_{fx}c^2_{x}c_c^2$. Eq. (\ref{reglike}) is an extension of the Regge-like formulas in Refs. \cite{Veseli:1996gy,Chen:2014nyo,Afonin:2014nya,Afonin:2020bqc,Chen:2017fcs}.

If the confining potential is $V(r)={\sigma}r^{a}$ ($a > 0$), Eq. (\ref{massform}) becomes
\bea\label{regv}
M=m_R+c_{fx}c_xc_c(x+c_{0x})^{2a/(a+2)}.
\eea
$c_x$ and $c_c$ are
\begin{align}\label{eq11}
c_l&=\left(1+\frac{a}{2}\right)
\left(\frac{1}{a}\right)^{a/(a+2)},\nonumber\\
c_{n_r}&=\left(\frac{1}{2}\right)^{a/(a+2)}
\left(\frac{a\pi}{B(1/a,3/2)}\right)^{2a/(a+2)},\nonumber\\
c_c&=\left(\frac{\sigma'^2}{\mu^{a}}\right)^{1/(a+2)},
\end{align}
where $B(x,y)$ denotes the beta function \cite{Gradshteyn:book1980}.
Correspondingly, Eq. (\ref{reglike}) becomes
\bea
(M-m_R)^2=\alpha_x(x+c_{0x})^{4a/(a+2)}.
\eea
For the heavy-heavy systems, the {\rts} plotted in the $\left((M-m_R)^2,\,x\right)$ plane will be convex upwards if $a>2/3$,  linear if $a=2/3$ and concave downwards if $a<2/3$ \cite{Chen:2018bbr}.

\section{Universal description of the heavy-heavy systems}\label{sec:fit}

In this section, we investigate the universality of the {\rt} relation (\ref{massform}) or (\ref{reglike}) by fitting the experimental and theoretical data. The {\rts} for the heavy-heavy systems are fitted individually by employing the formula (\ref{massform}). In this section, $c_{fx}$ and $c_{0x}$ are free parameters and vary with different {\rts} while other parameters are fixed for all heavy-heavy systems. To shown explicitly the nonlinearity of the {\rts} for the heavy-heavy systems, the fitted {\rts} are plotted in the $\left((M-m_R)^2,l\right)$ plane or in the $\left((M-m_R)^2,n_r\right)$ plane.

\subsection{Parameters}
The used parameters are   \cite{Feng:2023txx,Ebert:2011jc,Faustov:2021qqf,Faustov:2021hjs}
\begin{align}\label{paramet}
&m_c=1.55\, {\gev},\quad m_b=4.88\, {\gev},\quad \sigma=0.18\,{\gev^2},\nonumber\\
&C=-0.3\,{\gev},\quad m_{\{cc\}}=3.226\,{\gev}, \nonumber\\ &m_{\{bc\}}=6.526\,{\gev},\quad m_{\{bb\}}=9.778\,{\gev}, \nonumber\\
&m_{\{cu\}}=2.036\,{\gev},\quad m_{\{cs\}}=2.158\,{\gev}, \nonumber\\
&m_{\{bu\}}=5.381\,{\gev},\quad m_{\{bs\}}=5.482\,{\gev}.
\end{align}
$\{\,\}$ denotes the axial-vector diquark.
By using (\ref{massform}) with (\ref{parabm}) to fit the {\rts} individually, the parameters $c_{fl}$ (or $c_{fn_r}$) and $c_{0l}$ (or $c_{0n_r}$) are obtained, which are shown in figures.
The quality of a fit is measured by the quantity $\chi^2$ defined by \cite{Sonnenschein:2014jwa}
\bea\label{chi2}
\chi^2=\frac{1}{N-1}\sum_{i=1}^N\left(\frac{M_{fi}-M_{ei}}{M_{ei}}\right)^2,
\eea
where $N$ is the number of points on the trajectory, $M_{fi}$ is the fitted value and $M_{ei}$ is the experimental or theoretical value of the $i$-th particle.

The values in Eq. (\ref{paramet}) are used to calculate the masses of mesons \cite{Ebert:2011jc}, baryons \cite{Faustov:2021qqf} and tetraquarks \cite{Faustov:2021hjs}. These values are used to give an universal description of the heavy-light diquarks, mesons, baryons and tetraquarks \cite{Chen:2023web,Chen:2017fcs}. The values are also used to discuss diquarks \cite{Feng:2023txx,Chen:2023cws}.

\subsection{Doubly heavy diquarks}
In Ref. \cite{Feng:2023txx}, we show that the {\rts} for the doubly heavy diquarks can be well described by the (\ref{massform}) with (\ref{parabm}). The spectra of the doubly heavy diquarks $(cc)$, $(bc)$, and $(bb)$ obtained by using the {\rt} approach agree with other theoretical predictions.

\subsection{Heavy quarkonia}
In Ref. \cite{Chen:2018hnx}, we demonstrate that the spectra of the heavy quarkonia can be well described by the {\rt} relation (\ref{ortr}). Because
$\beta_x(x+c_{0x})^{2/3}{\ll}m_R$, (\ref{ortr}) can be obtained from (\ref{massform}) by neglecting the small term $(x+c_{0x})^{4/3}$. In the present work, we fit once again the {\rts} for the heavy quarkonia by using the relation (\ref{massform}).

\begin{table}[!phtb]
\caption{The experimental values (PDG) \cite{ParticleDataGroup:2022pth} and the theoretical values (EFG) \cite{Ebert:2011jc} for the radially and orbitally excited charmonia and bottomonia. The values are in {\gev}. $n$ is the radial quantum numbers plus one. $S$ is the total spin of the quark and antiquark. $J$ is the spin of the state, $P$ is the parity and $C$ is the charge parity.}  \label{tab:mesonr}
\centering
\begin{tabular*}{0.47\textwidth}{@{\extracolsep{\fill}}llll@{}}
\hline\hline
$n^{2S+1}$ $(J^{PC})$         &    Meson         & PDG    & EFG       \\
\hline
$1^3S_1(1^{--})$       &$J/\psi(1S)$  & 3.0969  & 3.096       \\
$2^3S_1(1^{--})$       & $\psi(2S)$   & 3.6861  & 3.685     \\
$3^3S_1(1^{--})$       & $\psi(4040)$  & 4.039   & 4.039       \\
$4^3S_1(1^{--})$       & $\psi(4415)$  & 4.421   & 4.427        \\
$5^3S_1(1^{--})$       &              &         & 4.837             \\
$6^3S_1(1^{--})$       &              &         & 5.167        \\
$1^3P_2(2^{++})$       & $\chi_{c2}(1P)$ & 3.55617   & 3.555   \\
$1^3D_3(3^{--})$       & $\psi_3(3842)$  & 3.84271   & 3.813  \\
$1^3F_4(4^{++})$       &                 &           & 4.093  \\
$1^3G_5(5^{--})$       &                  &           & 4.357       \\
$1^3H_6(6^{++})$       &                  &           & 4.608     \\
\hline\hline
State$(J^{PC})$         &    Meson         & PDG    & EFG        \\
\hline
$1^3S_1(1^{--})$       &$\Upsilon(1S)$  & $9.46040$    & 9.460         \\
$2^3S_1(1^{--})$       & $\Upsilon(2S)$ & $10.0234$   & 10.023      \\

$3^3S_1(1^{--})$       & $\Upsilon(3S)$ & $10.3551$    & 10.355       \\

$4^3S_1(1^{--})$       & $\Upsilon(4S)$ &  $10.5794$   & 10.586     \\

$5^3S_1(1^{--})$       & $\Upsilon(10860)$ & $10.8852$ & 10.869        \\
$6^3S_1(1^{--})$       & $\Upsilon(11020)$   &$11.000$  &11.088       \\
$1^3P_2(2^{++})$       & $\chi_{b2}(1P)$ & 9.91221  &9.912     \\
$1^3D_3(3^{--})$       &                 &          &10.166       \\
$1^3F_4(4^{++})$       &                 &          &10.349     \\
$1^3G_5(5^{--})$       &                  &         &10.514         \\
$1^3H_6(6^{++})$       &                  &         &10.672      \\
\hline\hline
\end{tabular*}
\end{table}

\begin{figure}[!phtb]
\centering
\includegraphics[scale=0.7]{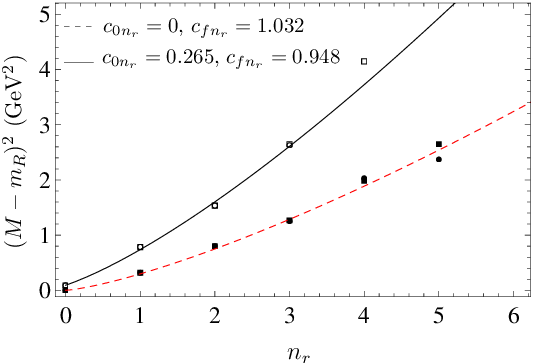}
\includegraphics[scale=0.7]{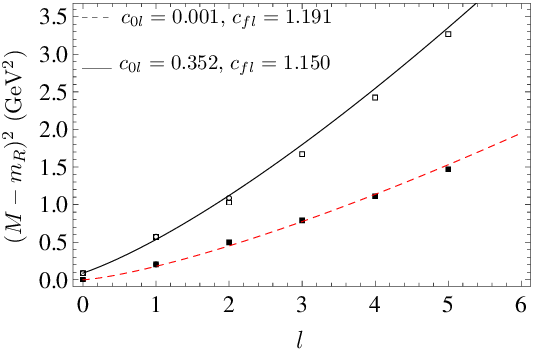}
\caption{The radial and orbital {\rts} for $J/\psi(1S)$ (the black line) and for $\Upsilon(1S)$ (the red dashed line), respectively. The experimental data (the circles and the filled circles) and the theoretical values (the empty squares and the filled squares) are listed in Table \ref{tab:mesonr}. (For interpretation of the colors in the figure(s), the reader is referred to the web version of this article.) }\label{fig:meson}
\end{figure}

As examples of the fully heavy mesons, the excited states of $J/\psi(1S)$ and $\Upsilon(1S)$ are discussed. The experimental and theoretical data are listed in Table \ref{tab:mesonr}. Eq. (\ref{massform}) is applied to fit the experimental data and the parameters $c_{fx}$ and $c_{0x}$ are calculated. $c_{fl}$ and $c_{0l}$ of the orbital {\rt} for the bottomonia are obtained by using the theoretical data. The obtained radial and orbital nonlinear {\rts} are plotted in the $\left((M-m_R)^2,\,n_r\right)$ plane and in the $\left((M-m_R)^2,\,l\right)$ plane, respectively. They are shown in Fig. \ref{fig:meson}. The experimental and theoretical data show that these radial and orbital {\rts} are not linear but obviously convex upwards.

According to Eqs. (\ref{massform}) and (\ref{cxcons}), $c_c$ will becomes smaller as the reduced mass $\mu$ increases. That explains why the radial (orbital) $\Upsilon(1S)$ {\rt} lies under the radial (orbital) $J/\psi(1S)$ {\rt}, see Fig. \ref{fig:meson}. As shown in Fig. \ref{fig:meson}, the parameters $c_{fn_r}$ and $c_{fl}$ for both the $J/\psi(1S)$ {\rt} and the $\Upsilon(1S)$ {\rt} are close to 1.

As shown in the upper panel of Fig. \ref{fig:meson}, the radial {\rts} for both $J/\psi(1S)$ and $\Upsilon(1S)$ agree well with the experimental data and with the theoretical data. This indicates that these mesons on the {\rts} satisfy Eq. (\ref{massform}), implying that these mesons can be described universally by Eq. (\ref{massform}).

The orbital {\rts} for the $J/\psi(1S)$ and $\Upsilon(1S)$ are shown in the lower panel of Fig. \ref{fig:meson}. The orbital {\rt} for the $J/\psi(1S)$ is in accordance with the experimental data and the theoretical data. The orbital $\Upsilon(1S)$ {\rt} is fitted by the theoretical data due to the limited numbers of experimental values. We can conclude that the orbitally excited states of $J/\psi(1S)$ and $\Upsilon(1S)$ satisfy Eq. (\ref{massform}), indicating an universal description.

\subsection{Heavy-heavy baryons composed of a heavy quark and a (doubly) heavy diquark}

\begin{table}[!phtb]
\caption{The theoretical values of the radially and orbitally excited states of the $\Omega_{ccb}$ $(3/2)^+$ state. The values are in {\gev}. $NL$ denote quantum numbers of the quark-diquark system. $N$ is the radial quantum numbers plus one. $L$ is the orbital quantum numbers. $J$ is the spin of the baryon and $P$ is the parity of the baryon.}  \label{tab:br}
\centering
\begin{tabular*}{0.47\textwidth}{@{\extracolsep{\fill}}ccccc@{}}
\hline\hline
$NL(J^P)$     &   FG \cite{Faustov:2021qqf}       & SGMG \cite{Serafin:2018aih}   & B \cite{Silvestre-Brac:1996myf}    & RP \cite{roberts:2008}    \\
\hline
$1S(3/2)^+$     & 7.999       & 8.301   & 8.056  & 8.265   \\
$2S(3/2)^+$     & 8.412       & 8.647   & 8.465  &    \\
$1P(5/2)^-$     & 8.267       & 8.491  & 8.331   &  8.432  \\
$1D(7/2)^+$     & 8.473       & 8.647   & 8.528   & 8.568   \\
\hline\hline
\end{tabular*}
\end{table}

\begin{table}[!phtb]
\caption{The theoretical values of the orbitally excited states of the $(1/2)^{-}$ states of $\Omega_{ccc}$, $\Omega_{bbb}$, and $\Omega_{cbb}$, respectively \cite{Faustov:2021qqf}. The values are in {\gev}. $NL$ denote quantum numbers of the quark-diquark system. $N$ is the radial quantum numbers plus one. $L$ is the orbital quantum numbers. $J$ is the spin of the state and $P$ is the parity.}  \label{tab:bebert}
\centering
\begin{tabular*}{0.47\textwidth}{@{\extracolsep{\fill}}cccc@{}}
\hline\hline
$NL(J^P)$     & $\Omega_{ccc}$ & $\Omega_{bbb}$ & $\Omega_{cbb}$   \\
\hline
$1P(1/2)^-$     & 5.010  & 14.698  & 11.414  \\
$1D(3/2)^+$     & 5.277  & 14.893  & 11.797  \\
$1F(5/2)^-$     & 5.519  & 15.081  &   \\
\hline\hline
\end{tabular*}
\end{table}

\begin{figure*}[!phtb]
\centering
\includegraphics[scale=0.45]{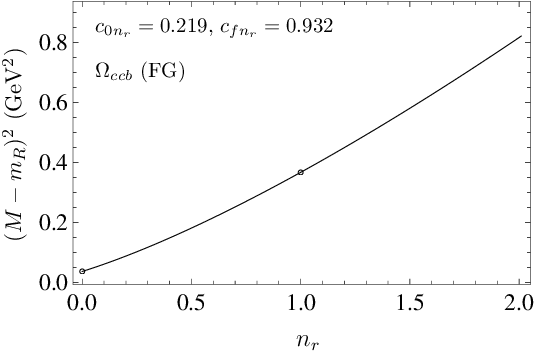}
\includegraphics[scale=0.45]{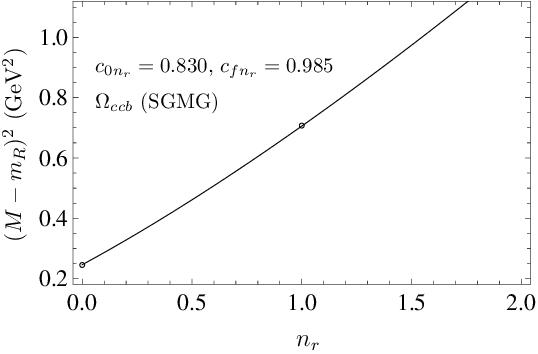}
\includegraphics[scale=0.45]{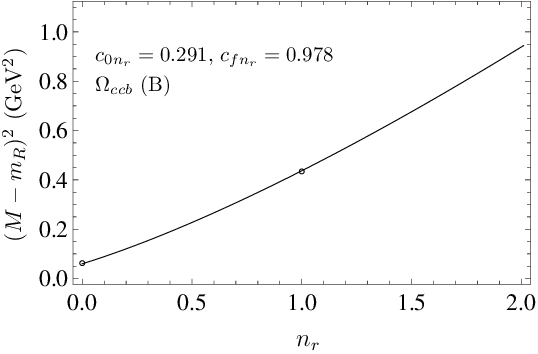}
\includegraphics[scale=0.45]{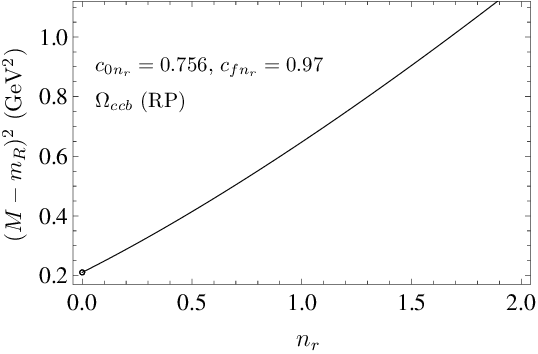}
\includegraphics[scale=0.45]{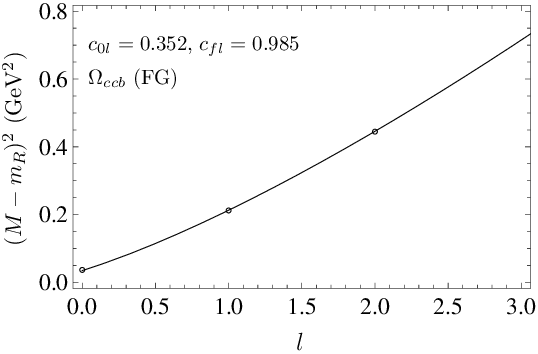}
\includegraphics[scale=0.45]{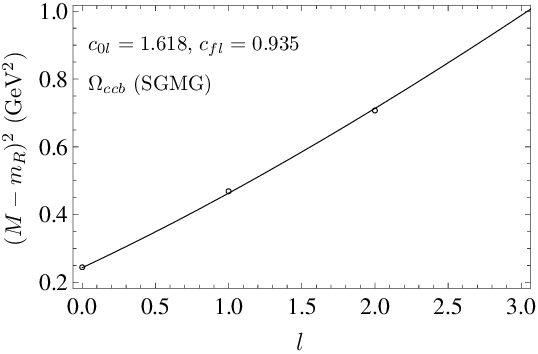}
\includegraphics[scale=0.45]{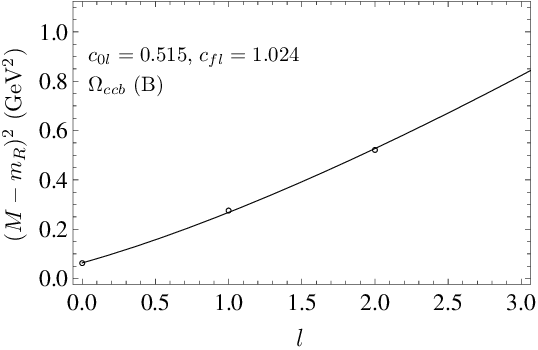}
\includegraphics[scale=0.45]{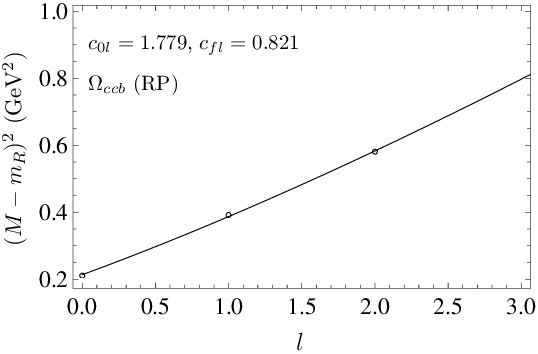}
\caption{The radial and orbital $\Omega_{ccb}$ {\rts}. The theoretical data are listed in Table \ref{tab:br}.  }\label{fig:br}
\end{figure*}

\begin{figure*}[!phtb]
\centering
\includegraphics[scale=0.6]{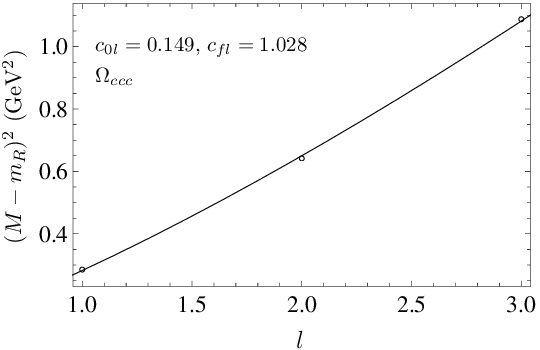}
\includegraphics[scale=0.6]{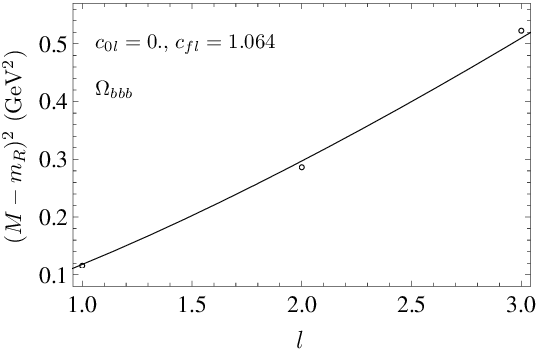}
\includegraphics[scale=0.6]{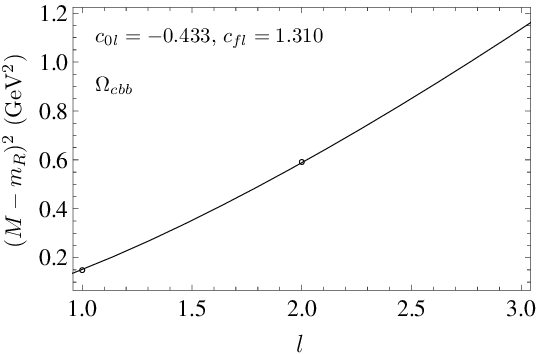}
\caption{The orbital {\rts} for $\Omega_{ccc}$, $\Omega_{bbb}$, and $\Omega_{cbb}$, respectively. The theoretical data are listed in Table \ref{tab:bebert}.  }\label{fig:bbr}
\end{figure*}

In the diquark picture \cite{Jaffe:2004ph,Selem:2006nd}, the heavy-heavy baryons consist of one heavy quark and one doubly heavy diquark or one heavy-light diquark. The radial and orbital baryon {\rts} for the $\Omega_{ccb}$ $(3/2)^+$ state and the orbital {\rts} for the $(1/2)^{-}$ states of $\Omega_{ccc}$, $\Omega_{bbb}$, and $\Omega_{cbb}$ are taken as examples to demonstrate the universality of the {\rt} relation (\ref{massform}).

The {\rts} are fitted by using the theoretical data for different models. The parameters $c_{fn_r}$, $c_{fl}$, $c_{0l}$ and $c_{0n_r}$ are determined by using the theoretical data.
For the RP model, $c_{fn_r}$ is set to be 0.97 due to the availability of only one theoretical value, and then $c_{0n_r}$ is calculated by using the theoretical data.
The data are listed in Tables \ref{tab:br} and \ref{tab:bebert}. The fitted radial and orbital {\rts} are shown in Figs. \ref{fig:br} and \ref{fig:bbr}.

From Figs. \ref{fig:br} and {\ref{fig:bbr}}, it can be observed that the parameters $c_{fn_r}$ and $c_{fl}$ for different models are close to 1.
The {\rts} for the baryons $\Omega_{ccb}$, $\Omega_{ccc}$, $\Omega_{bbb}$, and $\Omega_{cbb}$ satisfy Eq. (\ref{massform}), indicating these baryon states can be described universally by Eq. (\ref{massform}).

\subsection{Heavy-heavy tetraquarks composed of two heavy clusters}

\begin{table}[!phtb]
\caption{The theoretical values of $2^{++}$ states of the tetraquark $cc\bar{c}\bar{c}$. The values are in {\gev}. $NL$ denote quantum numbers of the diquark-antidiquark system. $N$ is the radial quantum numbers plus one. $L$ is the orbital quantum numbers. $J$ is the spin of the state, $P$ is the parity and $C$ is the charge parity.}  \label{tab:tr}
\centering
\begin{tabular*}{0.47\textwidth}{@{\extracolsep{\fill}}ccccc@{}}
\hline\hline
$NL(J^{PC})$     & FGS \cite{Faustov:2021hjs}        & DN \cite{Debastiani:2017msn}   & WLCLZ \cite{Wu:2016vtq}     & BFRS \cite{Bedolla:2019zwg}    \\
\hline
$1S(2^{++})$       & 6.367      & 6.1154   &6.194    & 6.246     \\
$2S(2^{++})$       & 6.868      & 6.6981   &         & 6.739    \\
$3S(2^{++})$       & 7.333      &          &         & 7.071      \\
$1P(3^{--})$     & 6.664      &6.6412   &    &     \\
$1D(4^{++})$     & 6.945      &         &    &       \\
\hline\hline
\end{tabular*}
\end{table}

\begin{table*}[!phtb]
\caption{The theoretical values of the radially excited $0^{++}$ states of the heavy-heavy tetraquarks \cite{Faustov:2022mvs,Faustov:2021hjs}. The values are in {\gev}. $NL$ denote quantum numbers of the diquark-antidiquark system. $N$ is the radial quantum numbers plus one. $L$ is the orbital quantum numbers.}  \label{tab:tall}
\centering
\begin{tabular*}{1.0\textwidth}{@{\extracolsep{\fill}}cccccccc@{}}
\hline\hline
$NL(J^{PC})$    & $cc\bar{c}\bar{c}$  & $cb\bar{c}\bar{b}$  & $bb\bar{b}\bar{b}$ & $cu\bar{c}\bar{u}$ & $cs\bar{c}\bar{s}$ & $bu\bar{b}\bar{u}$ & $bs\bar{b}\bar{s}$  \\
\hline
$1S(0^{++})$       &6.190   & 12.838  &19.315 &3.852 &4.110 &10.473 &10.671   \\
$2S(0^{++})$       &6.782   & 13.247  &19.680 &4.434 &4.680 &10.942 &11.133 \\
$3S(0^{++})$       &7.259   & 13.558  &19.941 & & & &   \\
\hline\hline
\end{tabular*}
\end{table*}

\begin{figure*}[!phtb]
\centering
\includegraphics[scale=0.45]{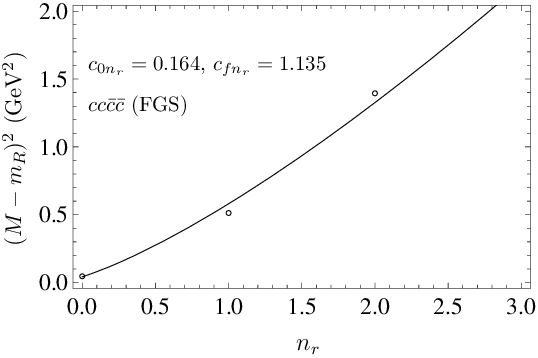}
\includegraphics[scale=0.45]{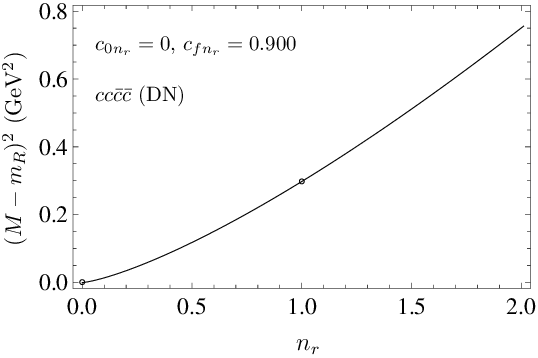}
\includegraphics[scale=0.45]{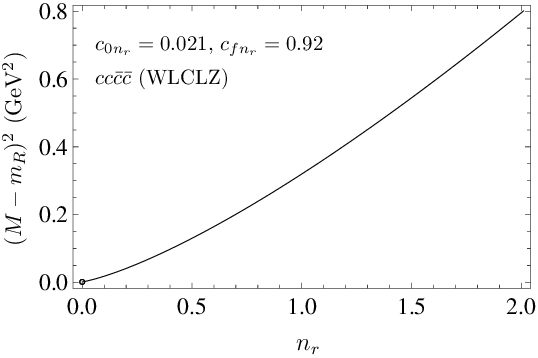}
\includegraphics[scale=0.45]{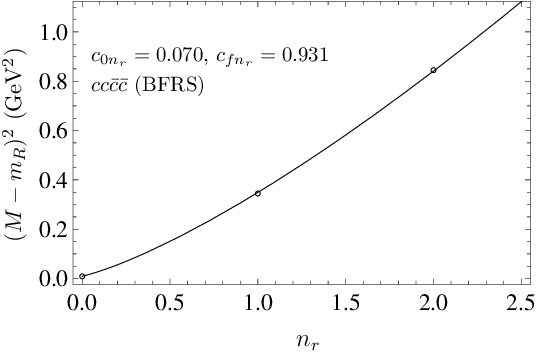}
\includegraphics[scale=0.45]{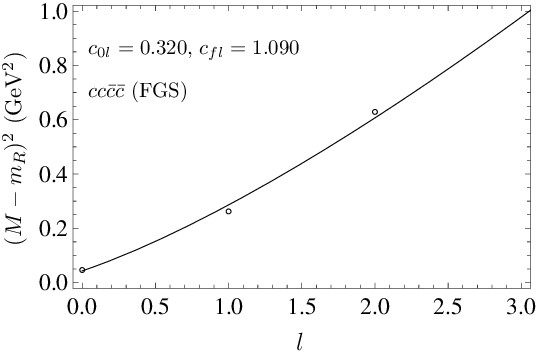}
\includegraphics[scale=0.45]{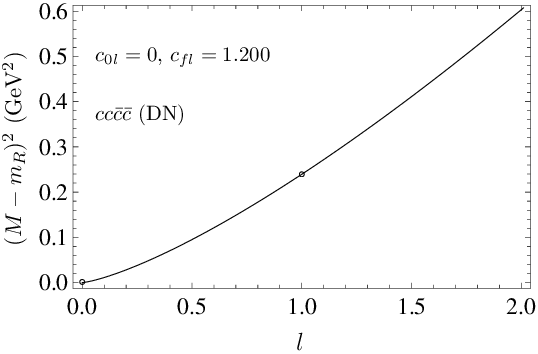}
\includegraphics[scale=0.45]{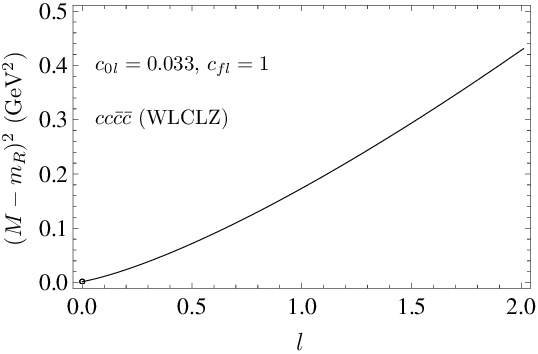}
\includegraphics[scale=0.45]{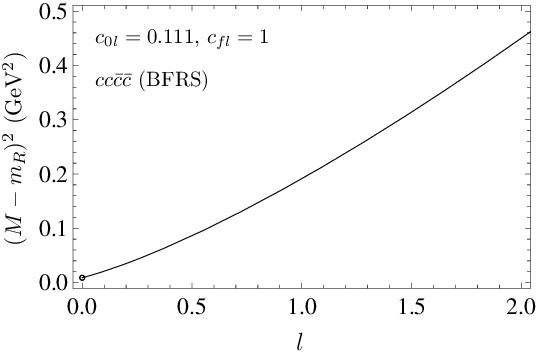}
\caption{The radial and orbital {\rts} for the $cc\bar{c}\bar{c}$ $2^{++}$ state in different models. The theoretical data are listed in Table \ref{tab:tr}.  }\label{fig:tr}
\end{figure*}

\begin{figure*}[!phtb]
\centering
\includegraphics[scale=0.45]{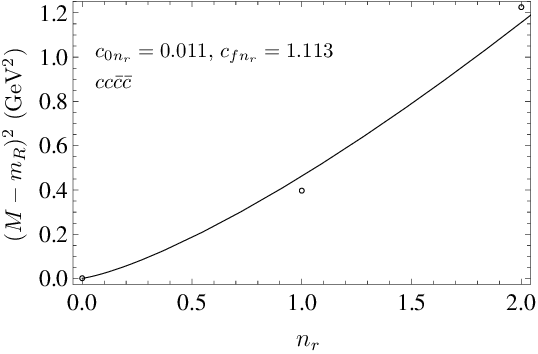}
\includegraphics[scale=0.45]{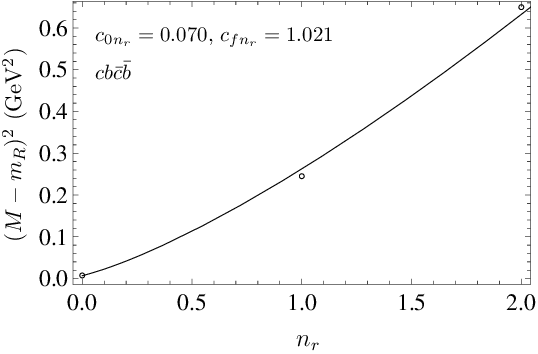}
\includegraphics[scale=0.45]{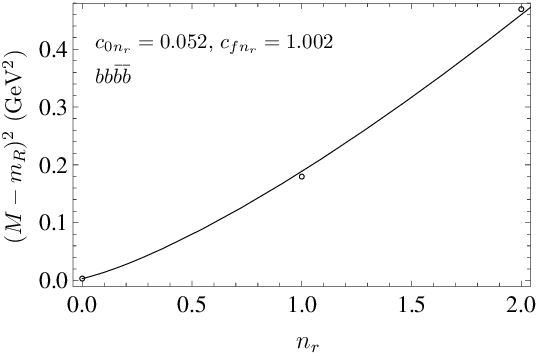}
\includegraphics[scale=0.45]{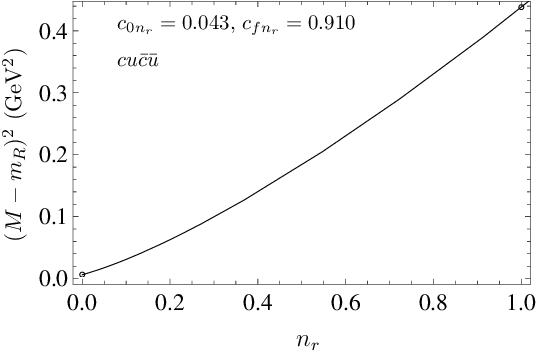}
\includegraphics[scale=0.45]{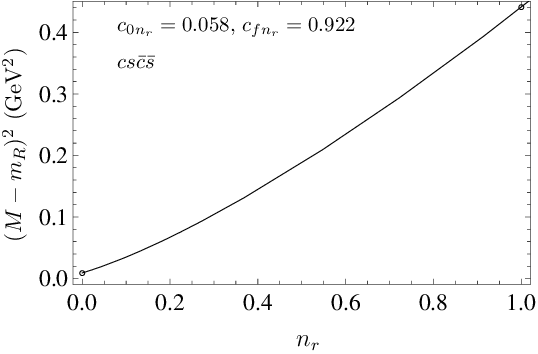}
\includegraphics[scale=0.45]{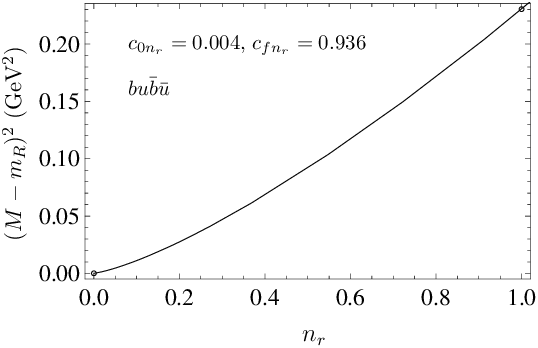}
\includegraphics[scale=0.45]{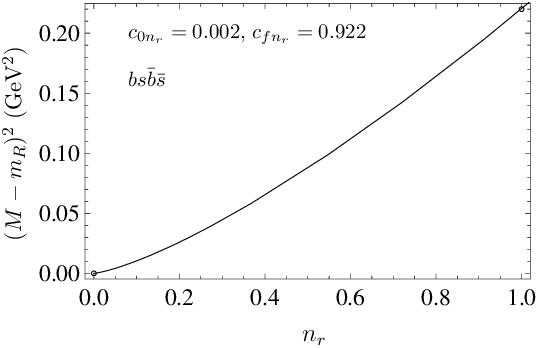}
\caption{The radial {\rts} for the $0^{++}$ states of $cc\bar{c}\bar{c}$, $cb\bar{c}\bar{b}$, $bb\bar{b}\bar{b}$, $cu\bar{c}\bar{u}$, $cs\bar{c}\bar{s}$, $bu\bar{b}\bar{u}$, and $bs\bar{b}\bar{s}$, respectively. The theoretical data are listed in Table \ref{tab:tall}.  }\label{fig:ttr}
\end{figure*}

In the diquark picture, the tetraquarks composed of one doubly heavy diquark or one heavy-light diquark and one doubly heavy antidiquark or one heavy-light antidiquark are the heavy-heavy systems. The
tetraquarks composed of a diquark and an antidiquark
in color $\bar{3}$ and $3$ configurations are considered.

The radial and orbital {\rts} for the $2^{++}$ state of the tetraquark ${cc\bar{c}\bar{c}}$ are fitted by employing Eq. (\ref{massform}). The theoretical data are calculated by different models. As the number of the points on a {\rt} are equal to or greater than two, the parameters $c_{fx}$ and $c_{0x}$ are fitted while $c_{fx}$ is set to a value if there is only one point on a {\rt}. The theoretical data are listed in Table \ref{tab:tr}. The fitted {\rts} are shown in Fig. \ref{fig:tr}.

As seen in Fig. \ref{fig:tr}, the parameters $c_{fn_r}$ and $c_{fl}$ for different models are close to 1.
Eq. (\ref{massform}) is satisfied with the radially excited states and the orbitally excited states of the $2^{++}$ state of the tetraquark ${cc\bar{c}\bar{c}}$.
The fitted {\rts} are similar to each other, demonstrating the consistency of these models and suggesting that these models are close to experiments. Therefore, we conclude that these excited states of the $2^{++}$ state of the tetraquark ${cc\bar{c}\bar{c}}$ can be described universally by Eq. (\ref{massform}), despite the lack of experimental data. It is expected that other states of tetraquarks will satisfy the universal relation (\ref{massform}). As a check, we fit the radial {\rts} for the $cc\bar{c}\bar{c}$, $cb\bar{c}\bar{b}$, $bb\bar{b}\bar{b}$, $cq\bar{c}\bar{q}$, $cs\bar{c}\bar{s}$, $bq\bar{b}\bar{q}$, and $bs\bar{b}\bar{s}$, respectively, see Fig. \ref{fig:ttr} and Table \ref{tab:tall}. The universality of relation (\ref{massform}) is confirmed once again.

\section{Discussions}\label{sec:disc}

As discussed in the previous section, all heavy-heavy systems, including the doubly heavy diquarks, the heavy quarkonia, the heavy-heavy baryons and the heavy-heavy tetraquarks, can be described universally by the relation (\ref{massform}). The universality encompasses not only the universal behavior $M{\sim}x^{2/3}$ but also the universal parameters. $c_{fn_r}$ and $c_{fl}$ vary with different heavy-heavy systems and they are close to one. There is an inequality $\beta_{n_r}>\beta_{l}$ and it holds for all the discussed heavy-heavy systems. Moreover, the expression of $\beta_x$ [Eq. (\ref{parabm})] explains its variation with change of the constituent's masses.

In potential models, the form of the {\rt} relation is determined by the dynamic equation. Different forms of kinematic terms corresponding to different energy regions will yield different behaviors of the {\rts} \cite{Chen:2021kfw,Chen:2022flh,Chen:2018bbr}. ${\bf p}$ and $r^a$ leads to $M{\sim}x^{a/(a+1)}$ while ${\bf p}^2$ and $r^a$ gives $M{\sim}x^{2a/(a+2)}$ $(x=l,\,n_r)$. The discussions in the previous section show that the {\rts} for the heavy-heavy systems behave as $M{\sim}x^{2/3}$. It is consistent with the fact that the heavy constituents in the heavy-heavy systems are nonrelativistic.

That the fitted {\rts} for the heavy-heavy systems behave as $M{\sim}x^{2/3}$ shows that the confining potential in these heavy-heavy systems is linear according to Eq. (\ref{regv}). The confining potential plays a dominant role in the {\rts} for the heavy-heavy systems. The effect of the color Coulomb potential becomes important when considering the spin-dependent terms.

\begin{table}[!phtb]
\caption{The experimental and theoretical values of the radially and orbitally excited states of $B_c$, respectively. The values are in {\gev}. $n$ is the radial quantum numbers plus one. $S$ is the total spin of the quark and antiquark. $J$ is the spin of the state, $P$ is the parity.}  \label{tab:mesonbc}
\centering
\begin{tabular*}{0.47\textwidth}{@{\extracolsep{\fill}}cccccc@{}}
\hline\hline
$n^{2S+1}$ $(J^P)$         &    Meson         & PDG \cite{ParticleDataGroup:2022pth}    & Fitted           & EFG \cite{Ebert:2011jc}  & G \cite{Godfrey:2004ya}  \\
\hline
$1^1S_0\,(0^-)$       &$B_c^{\pm}(1S)$  & 6.27447  & Input & 6.272 & 6.271     \\
$2^1S_0\,(0^-)$       & $B_c(2S)^{\pm}$ & 6.8712    & 6.86  & 6.842 & 6.855  \\
$3^1S_0\,(0^-)$       &                 &           & 7.25  & 7.226 & 7.250    \\
$4^1S_0\,(0^-)$       &                 &           & 7.59  & 7.585 &     \\
$5^1S_0\,(0^-)$       &                 &           & 7.88  & 7.928 &  \\
$6^1S_0\,(0^-)$       &                 &           & 8.16  &      & \\
$1^1P_1\,(1^+)$       &                 &          & 6.71  & 6.750 & 6.750 \\
$1^1D_2\,(2^-)$       &                 &           & 7.01  & 7.026 & 7.036 \\
$1^1F_3\,(3^+)$       &                 &           & 7.27  & 7.268 & 7.266 \\
$1^1G_4\,(4^-)$       &                 &           & 7.50  & 7.487  \\
$1^1H_5\,(5^+)$       &                 &           & 7.71  &   \\
\hline\hline
\end{tabular*}
\end{table}

The {\rts} for the heavy-heavy systems are not only nonlinear in the $(M,\,x)$ plane but also nonlinear in the $(M^2,\,x)$ plane \cite{Chen:2018hnx} and  in the $((M-m_R)^2,\,x)$ plane.
From Eq. (\ref{massform}), we have
\bea\label{msforms}
M^2=m_R^2+\beta_x^2(x+c_{0x})^{4/3}+2m_R\beta_x(x+c_{0x})^{2/3}.
\eea
Eq. (\ref{msforms}) behaves as $M^2{\sim}x^{2/3}$ as $m_R{\gg}\beta_x(x+c_{0x})^{2/3}$. It can be approximately linear $M^2{\sim}x$ as $m_R$ is roughly equal to $\beta_x(x+c_{0x})^{2/3}$. In the $(M^2,\,x)$ plane, the nonlinearity of the {\rt} for the heavy-heavy systems becomes more intense as $m_R$ becomes larger. Therefore, the nonlinearity of the {\rts} for bottomonia is more intense than that for charmonia. In the $((M-m_R)^2,\,x)$ plane, the nonlinearity is obvious as $m_R$ is chosen appropriately \cite{Chen:2022flh}, see the figures in the present work.

As a check of the universality of relation (\ref{massform}), we use the relation to predict the masses of the $B_c$ mesons. As shown in Fig. \ref{fig:meson}, $c_{fn_r}$ and $c_{fl}$ vary with the constituents' masses. We use the theoretical data in Ref. \cite{Ebert:2011jc} to fit the radial and orbital {\rts} for the $1^1S_0$ states of the charmonia and bottomonia, respectively. We choose $c_{fn_r}=1.017$, $c_{fl}=1.174$ by averaging the values of $c_{fn_r}$ and $c_{fl}$ for the charmonia and bottomonia because the experimental data of $B_c$ are scarce. Using the parameters in (\ref{paramet}), Eq. (\ref{massform}) and the experimental data of $B_c^{\pm}(1S)$, we obtain $c_{0n_r}=0.097$ and $c_{0l}=0.142$. After all parameters are determined, the fitted results are calculated and they are in agreement with other theoretical results, see Table \ref{tab:mesonbc}. The mixing of the singlet state and the triplet state are not considered. The $1^1l_l$ states are calculated by using the orbital {\rt} for the $1^1S_0$ state.

\section{Conclusions}\label{sec:conc}
We employ the nonlinear {\rt} relation $M=m_R+\beta_x(x+c_{0x})^{2/3}$ $(x=l,\,n_r)$ to fit the {\rts} for the doubly heavy diquarks, the fully heavy mesons, the baryons composed of one heavy quark and one doubly heavy diquark and the tetraquarks consisting of two heavy clusters. The fitted results illustrate the universal description of these heavy-heavy systems. The {\rt} relation is expected to be applied to other multiquarks made up of two heavy clusters.

The universality embodies not only the universal behavior $M{\sim}x^{2/3}$ but also the universal parameters. The parameters $c_{fn_r}$ and $c_{fl}$ vary with different heavy-heavy systems and the values of them are all close to 1. As $c_{fn_r}$ and $c_{fl}$ are determined, the universal relation has only one free parameter $c_{0x}$ for one {\rt} which can be fitted by one point on the given {\rt}. Therefore, the universal relation can estimate all other states on a {\rt} only if one point is known.

There is an inequality $\beta_{n_r}>\beta_{l}$ and it holds for all the disussed heavy-heavy systems. In addition, the expression of $\beta_x$ [Eq. (\ref{parabm})] explains its variation with change of the constituents' masses.

\section*{Acknowledgments}
We are very grateful to the anonymous referees for the valuable comments and suggestions.

\end{document}